\documentstyle[twocolumn,prl,aps,epsfig,amsfonts]{revtex}

\begin{document}
\draft
\preprint{}

\newcommand{\1}{{\bf \scriptstyle 1}\!\!{1}}
\newcommand{\I}{{\rm i}}
\newcommand{\p}{\partial}
\newcommand{\D}{^{\dagger}}
\newcommand{\bx}{{\bf x}}
\newcommand{\bk}{{\bf k}}
\newcommand{\bv}{{\bf v}}
\newcommand{\bp}{{\bf p}}
\newcommand{\bu}{{\bf u}}
\newcommand{\bA}{{\bf A}}
\newcommand{\bB}{{\bf B}}
\newcommand{\bE}{{\bf E}}
\newcommand{\bF}{{\bf F}}
\newcommand{\bI}{{\bf I}}
\newcommand{\bK}{{\bf K}}
\newcommand{\bL}{{\bf L}}
\newcommand{\bP}{{\bf P}}
\newcommand{\bQ}{{\bf Q}}
\newcommand{\bS}{{\bf S}}
\newcommand{\bH}{{\bf H}}
\newcommand{\balpha}{\mbox{\boldmath $\alpha$}}
\newcommand{\bsigma}{\mbox{\boldmath $\sigma$}}
\newcommand{\bSigma}{\mbox{\boldmath $\Sigma$}}
\newcommand{\bOmega}{\mbox{\boldmath $\Omega$}}
\newcommand{\bpi}{\mbox{\boldmath $\pi$}}
\newcommand{\bphi}{\mbox{\boldmath $\phi$}}
\newcommand{\bnabla}{\mbox{\boldmath $\nabla$}}
\newcommand{\bmu}{\mbox{\boldmath $\mu$}}
\newcommand{\bepsilon}{\mbox{\boldmath $\epsilon$}}

\newcommand{\iLambda}{{\it \Lambda}}
\newcommand{\cA}{{\cal A}}
\newcommand{\cD}{{\cal D}}
\newcommand{\cF}{{\cal F}}
\newcommand{\cL}{{\cal L}}
\newcommand{\cH}{{\cal H}}
\newcommand{\cI}{{\cal I}}
\newcommand{\cM}{{\cal M}}
\newcommand{\cO}{{\cal O}}
\newcommand{\cR}{{\cal R}}
\newcommand{\cU}{{\cal U}}
\newcommand{\cT}{{\cal T}}

\newcommand{\be}{\begin{equation}}
\newcommand{\ee}{\end{equation}}
\newcommand{\bea}{\begin{eqnarray}}
\newcommand{\eea}{\end{eqnarray}}
\newcommand{\beqa}{\begin{eqnarray*}}
\newcommand{\eeqa}{\end{eqnarray*}}
\newcommand{\nn}{\nonumber}
\newcommand{\DD}{\displaystyle}

\newcommand{\ba}{\left[\begin{array}{c}}
\newcommand{\baa}{\left[\begin{array}{cc}}
\newcommand{\baaa}{\left[\begin{array}{ccc}}
\newcommand{\baaaa}{\left[\begin{array}{cccc}}
\newcommand{\ea}{\end{array}\right]}

\twocolumn[
\hsize\textwidth\columnwidth\hsize\csname
@twocolumnfalse\endcsname

\title{Measuring the entanglement of coupled spins by multiphoton interference}

\author{Michael N.~Leuenberger and Michael E. Flatt\'e
}
\address{Department of Physics and Astronomy, University of Iowa,
IATL, Iowa City, IA 52242, USA}
\author{D.~D.~Awschalom
}
\address{Department of Physics, University of California, Santa Barbara,
CA 93106-9530, USA}

\date{\today}
\maketitle

\begin{abstract}
We propose an experimental method to measure the entanglement of coupled spins, each in a separate quantum dot,
by means of multiphoton interference patterns generated through the scattering of two laser beams
off the quantum dots. We calculate the $N$-photon quantum correlations measured by $N$ detectors on an image plane. Using two perpendicular laser beams, either many correlation measurements on a time ensemble or a single correlation measurement on a spatial ensemble of the many-qubit state is sufficient to retrieve all the possible amplitudes of a many-qubit state.
\end{abstract}

]
\narrowtext

Semiconductor spintronics\cite{Wolf} and semiconductor quantum computation\cite{Awschalom} have become very popular research fields since 
time-resolved Faraday and Kerr rotation measurements in GaAs semiconductors revealed an electron spin coherence length and time exceeding 100 $\mu$m and  100 ns, respectively\cite{Kikkawa1998,Kikkawa1999}. Since then $T_2^*$'s exceeding 1 ns have been observed in undoped quantum dot ensembles\cite{Gupta1999}. Most of the quantum computation schemes for semiconductor spin systems can be traced back to the initial proposal in \cite{Loss}. Information on the result of a quantum computation, i.e. the exact many-qubit state of a spin system, was proposed to be measureable indirectly through the charge degree of freedom.
Of special interest is the entanglement of two or more spins, which can be measured for a non-degenerate eigenstate, for example, in the noise of the charge current flowing through two quantum dots\cite{Loss&Sukhorukov}, or through the production of  
single pairs of entangled photons\cite{Gywat}. 

Here we propose a general and robust method of measuring optically all the information about the entangled state permitted by quantum measurement bounds, independently of whether or not the eigenstates are non-degenerate. In the case of a spatial ensemble of spin systems coherent oscillations of many-spin states can be observed by repetitive measurements without the need to refresh the many-qubit state, analogous to the precession of an ensemble of single spins. Repetitive preparation and detection of the many-qubit state can also be performed. These proposed measurements for $n$ qubits are the $n$-qubit generalization of single-spin Faraday rotation\cite{Awschalom}. For the measurement of a single many-qubit state, repetitive measurement without preparation provides that information permitted by quantum measurement bounds.
Our approach reads out the many-qubit states of the spins of excess electrons in quantum dots by means of nonresonant laser beam scattering off the valence band electrons [via virtual excitation of electron-hole pairs (see Fig.~\ref{Entanglement_detection_interference})]. Such virtual states also play a central role in the optical Stark effect\cite{Gupta,Pryor}. Since the selection rule for the creation of the virtual electron-hole pairs matches exactly the spin state of the excess electron, the entanglement of the excess electrons is imprinted onto the many-photon system.
Two perpendicular laser beams with different circular polarizations illuminate the sample, and the amplitudes and the entanglement of the many-photon state are read out by multiphoton correlation interference patterns.

We first demonstrate our method for a two-qubit system. We start from a general two-spin state of the form
\be
\left|\psi\right>=a_{++}\left|\uparrow;\uparrow\right>+a_{+-}\left|\uparrow;\downarrow\right>+a_{-+}\left|\downarrow;\uparrow\right>
+a_{--}\left|\downarrow;\downarrow\right>,
\label{qubits}
\ee
where $\left|\uparrow\right>$ and $\left|\downarrow\right>$ are the spin states of the excess electrons in quantum dot 1 and 2, separated by a semicolon.  We are interested in reading out the amplitudes $a_{++},a_{+-},a_{-+},a_{--}$, which also determine the entanglement of the two spins. 


To measure the amplitudes two perpendicular and circularly polarized laser beams are scattered off the two quantum dots (see Fig.~\ref{Entanglement_detection_interference}). The electric component of the laser field can be expressed by
$
\bE(\bx,t)=\bE^{(+)}(\bx,t)+\bE^{(-)}(\bx,t),
$
where 
\be
\bE^{(\pm)}(\bx,t)=\pm i\sum_{\bk\sigma}\sqrt{\frac{\hbar\omega_k}{2\varepsilon_0V}}\bepsilon_{\bk\sigma}a_{\bk\sigma}^{(\dagger)}e^{\pm i(\bk\cdot\bx-\omega t)}.
\ee
$a_{\bk\sigma^\pm}$ and $a_{\bk\sigma^\pm}\D$ are the annihilation and creation operator of a $\sigma^\pm$ photon, respectively. $\varepsilon_0$ is the vacuum dielectric constant, and $V$ is the volume containing our quantized electric field $\bE(\bx,t)$. We restrict ourselves to one mode $\bk$ and one frequency $\omega_\bk$. For convenience, we choose the left and right circular polarizations $\sigma^\pm$ for our calculations.
We wish to avoid absorption of the laser light, so the frequency of the laser should be tuned slightly below the bandgap of the semiconductor. For, e.g., ZnSe dots, one could use a wavelength of $\lambda=500$ nm. The light induces Raman transitions of the valence electrons in the $J_z=\pm 3/2$ states (see Fig.~\ref{Entanglement_detection_levels}). The virtual states are marked by dashed lines in Fig.~\ref{Entanglement_detection_levels}.
The electron-photon interaction Hamiltonian reads\cite{Cardona}
$
\cH_{ep}=\sum_{i=1}^n\frac{e}{mc}\bA_i\cdot\bp_i,
$
where $\bA_i$ is the vector potential of the laser interacting with quantum dot $i$ with $\bE_i=\p\bA_i/\p t$, $\bp_i$ is the momentum of the electron with spin $J_z=\pm 3/2$ on quantum dot $i$, and $n$ is the number of quantum dots. The left/right circularly polarized laser beam propagates at 45$^\circ$/-45$^\circ$ angle with respect to the axis that is perpendicular to the image plane. In order for the two beams to be directed into the image plane, two mirrors at 22.5$^\circ$/-22.5$^\circ$ angle must be set up right next to the sample, which is shown in Fig.~\ref{Entanglement_detection_interference}. The quantum dots can be arranged in any 2D configuration, including the random distribution expected from growth kinetics. The rotationally invariant selection rules of spherical dots give the cleanest results. So we will focus only on them.

\begin{figure}[htb]
  \begin{center}
    \leavevmode
\epsfxsize=7cm
\epsffile{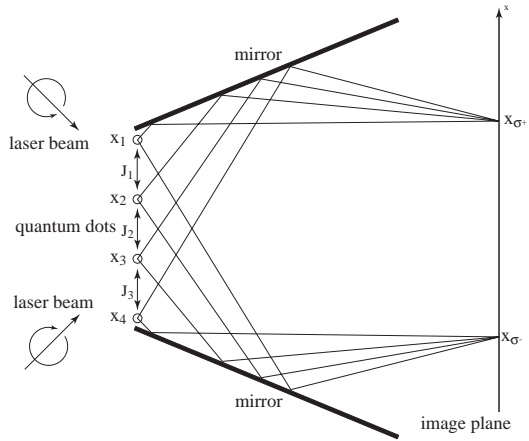}
  \end{center}
\caption{Detection scheme where both laser beams are scattered off the quantum dots, which produces
an interference pattern on the image plane. The left and right circularly polarized laser beams propagate at 45$^\circ$ and -45$^\circ$ angles. After scattering off the sample, both laser beams are redirected towards the image plane by two mirrors at 22.5$^\circ$ and -22.5$^\circ$ angles. The distance between both outgoing parallel beams can be made of the order of the sample size. Therefore the pattern that is seen on the image plane comes from almost the same source image.}
\label{Entanglement_detection_interference}
\end{figure}

It is important to note that there is only an exchange interaction between the excess electrons with spin $\pm 1/2$ of the quantum dots, and not between the valence electrons with spin $J_z=\pm 3/2$, due to the Pauli exclusion principle. Thus we can treat the electron-photon interactions for each quantum dot separately, i.e. the photons produce virtual electron-hole pairs on each quantum dot separately. Since $\left[\bA_i\cdot\bp_i,\bA_j\cdot\bp_j\right]=0$ for $i\ne j$, we know that the propagator splits like
$e^{\frac{i}{\hbar}\cH_{ep}}=\prod_{i=1}^ne^{\frac{i}{\hbar}\frac{e}{mc}\bA_i\cdot\bp_i}$.
Since our quantum dots are spherically symmetric, we can use any quantization axis we want.
For the calculation it is best to choose the quantization axis -- from now on our z-axis -- in the direction of one of the laser beams. We choose the $\sigma_{(z)}^-$ photons to propagate in z-direction, and the $\sigma_{(x)}^+$ photons to propagate in x-direction.
Then the qubits select the appropriate circular polarization according to their spin state, which is shown in Fig.~\ref{Entanglement_detection_levels}.
If the exchange coupling between the quantum dots is much larger than the electron-photon coupling, the one-photon scattering is suppressed, because the exchange is blocked only for a two-photon scattering, i.e. the energy levels of one- and two-photon absorption are different. The same holds for $n$-photon scattering, where $p$-photon scattering is suppressed for $0<p<n$. If the quantum dots do not interact, there will be spurious coincidences. Although this leads to a background intensity, these coincidences are random and thus are not seen in the correlation interference pattern. After taking the direct product of Eq.~(\ref{qubits}) and $\left|\sigma_{(x)}^+\right>+\left|\sigma_{(z)}^-\right>$, we obtain first a transition to the intermediate state
\bea
\left|\psi\right> & = & \tilde{a}_{++}\left|\leftarrow\underline{\rightarrow};\leftarrow\underline{\rightarrow}\right>
+\tilde{a}_{+-}\left|\leftarrow\underline{\rightarrow};\underline{\uparrow}\downarrow\right> \nn\\
& & +\tilde{a}_{-+}\left|\underline{\uparrow}\downarrow;\leftarrow\underline{\rightarrow}\right>
+\tilde{a}_{--}\left|\underline{\uparrow}\downarrow;\underline{\uparrow}\downarrow\right>,
\eea
where the underlined arrows indicate the virtual electron-hole pair and $\tilde{a}_{++}=\frac{1}{2}(a_{++}+a_{+-}+a_{-+}+a_{--})$, 
$\tilde{a}_{+-}=\frac{1}{\sqrt{2}}(a_{+-}+a_{--})$, $\tilde{a}_{-+}=\frac{1}{\sqrt{2}}(a_{-+}+a_{--})$, $\tilde{a}_{--}=a_{--}$. The next virtual transition takes the combined electron-photon state into
\bea
\left|\psi_{\rm ep}\right> & = & \tilde{a}_{++}\left|\rightarrow;\rightarrow\right>\left|\sigma_{(x)}^+;\sigma_{(x)}^+\right> 
+\tilde{a}_{+-}\left|\rightarrow;\downarrow\right>\left|\sigma_{(x)}^+;\sigma_{(z)}^-\right> \nn\\
& & +\tilde{a}_{-+}\left|\downarrow;\rightarrow\right>\left|\sigma_{(z)}^-;\sigma_{(x)}^+\right> 
+\tilde{a}_{--}\left|\downarrow;\downarrow\right>\left|\sigma_{(z)}^-;\sigma_{(z)}^-\right>,
\label{wavefunction}
\eea
where the semicolon for the photon states separates the photon coming from quantum dot 1 and 2, respectively.
Note that only a small part of the incoming laser beam gets entangled with the electrons (the usual source of decoherence\cite{Zhou}), which is determined by the scattering amplitude.
Even if the quantum dots have different energy level spacings, the amplitude distribution $a_{++},a_{+-},a_{-+},a_{--}$ is preserved, because the $\sigma^+$ and $\sigma^-$ dipole transition amplitudes are equal. 

\begin{figure}[htb]
  \begin{center}
    \leavevmode
\epsfxsize=7cm
\epsffile{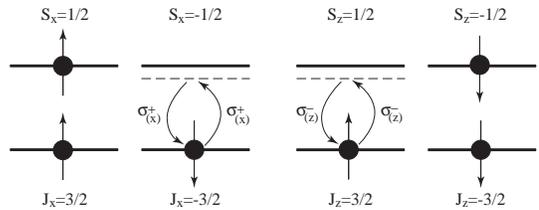}
  \end{center}
\caption{Energy level diagrams for a two-qubit state. The Raman transition is produced by a $\sigma_{(x)}^+$ photon when the excess electron is in the state $\left|\leftarrow\right>=(\left|\uparrow\right>+\left|\downarrow\right>)/\protect\sqrt{2}$, or by a $\sigma_{(z)}^-$ photon when the excess electron is in the state $\left|\downarrow\right>$.}
\label{Entanglement_detection_levels}
\end{figure}

Now we are ready to measure the two-photon correlations in order to retrieve the amplitudes of the two-qubit state. The joint probability that one photon with polarization $\sigma^+$ is detected at $x_{\sigma_+}$ and a second photon with polarization $\sigma^-$ is detected at $x_{\sigma_-}$ (see Fig.~\ref{Entanglement_detection_interference}) is proportional to
$\left<\bE_{\sigma^+}^{(-)}\bE_{\sigma^-}^{(-)}\bE_{\sigma^-}^{(+)}\bE_{\sigma^+}^{(+)}\right>$ \cite{Mandel}.
Each electric field term consists of a sum of electric fields coming from all the quantum dots, i.e.
$
\bE_{\sigma^\pm}^{(\pm)}=\sum_{i=1}^n \bE_{i,\sigma^\pm}^{(\pm)},
$
where $n$ is the number of quantum dots.
Then the two-photon interference pattern of two quantum dots is given by the two-photon correlation
\be
C_2=\left|\tilde{a}_{+-}\right|^2+\left|\tilde{a}_{-+}\right|^2+\left|\tilde{a}_{+-}\right|\left|\tilde{a}_{-+}\right|
\cos(\xi+\varphi),
\label{C_2}
\ee
where $\xi=k(\left|\bx_1-\bx_{\sigma^+}\right|+\left|\bx_2-\bx_{\sigma^-}\right|-\left|\bx_1-\bx_{\sigma^-}\right|-\left|\bx_2-\bx_{\sigma^+}\right|)$.
From this interference pattern we can retrieve $\left|\tilde{a}_{+-}\right|$, $\left|\tilde{a}_{-+}\right|$, and the phase difference 
$\varphi=\varphi_{+-}-\varphi_{-+}$ between $\tilde{a}_{+-}$ and $\tilde{a}_{-+}$, which is a direct measure of the entanglement regarding 
$\left|\uparrow;\downarrow\right>$ and $\left|\downarrow;\uparrow\right>$.
For example, for coherent oscillations between $(\left|\uparrow;\downarrow\right>-\left|\downarrow;\uparrow\right>)/\sqrt{2}$ and $(\left|\uparrow;\downarrow\right>+\left|\downarrow;\uparrow\right>)/\sqrt{2}$, $\sqrt{2}\tilde{a}_{+-}=a_{+-}$ and $\sqrt{2}\tilde{a}_{-+}=a_{-+}$.
Note that the two detectors must be locally separated since the entanglement is a nonlocal physical phenomenon.
In order to detect the entanglement regarding $\left|\uparrow;\uparrow\right>$ and $\left|\downarrow;\downarrow\right>$, 
the spins would have to be rotated globally into the direction
with spherical angles $\theta$ and $\phi$, which transform the spin states into
$\left|\uparrow'\right>=\cos\frac{\theta}{2}\left|\uparrow\right>+\sin\frac{\theta}{2}e^{i\phi}\left|\downarrow\right>$,
$\left|\downarrow'\right>=-\sin\frac{\theta}{2}\left|\uparrow\right>+\cos\frac{\theta}{2}e^{i\phi}\left|\downarrow\right>$.
The transformed two-qubit amplitudes read then
\bea
a_{++}' & = & a_{++}c^2-(a_{+-}+a_{-+})cs+a_{--}s^2, \\
a_{+-}' & = & [(a_{++}-a_{--})cs+a_{+-}c^2-a_{-+}s^2]e^{i\phi}, \\
a_{-+}' & = & [(a_{++}-a_{--})cs-a_{+-}s^2+a_{-+}c^2]e^{i\phi}, \\
a_{--}' & = & [a_{++}s^2+(a_{+-}+a_{-+})cs+a_{--}c^2]e^{2i\phi},
\eea
where $c=\cos\frac{\theta}{2}$, $s=\sin\frac{\theta}{2}$.

\begin{figure}[htb]
  \begin{center}
    \leavevmode
\epsfxsize=7cm
\epsffile{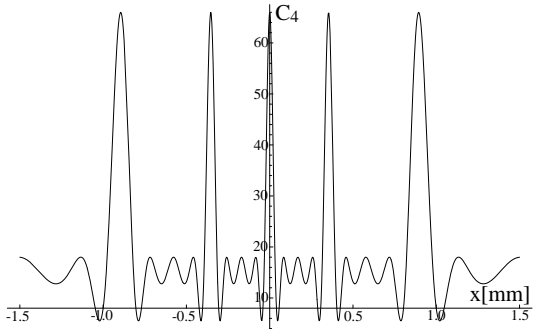}
  \end{center}
\caption{Here a 2-photon correlation interference pattern of a 4-qubit state has been calculated by means of Eq.~(\protect\ref{C_4}).}
\label{Entanglement_detection_pattern}
\end{figure}

Consequently, the correlation is transformed to
\bea
C_2'=\left|\tilde{a}'_{+-}\right|^2+\left|\tilde{a}'_{-+}\right|^2+\left|\tilde{a}'_{+-}\right|\left|\tilde{a}'_{-+}\right|
\cos(\xi+\varphi'), 
\eea
where $\tilde{a}'_{++}=\frac{1}{2}(a_{++}'+a_{+-}'+a_{-+}'+a_{--}')$, 
$\tilde{a}_{+-}'=\frac{1}{\sqrt{2}}(a_{+-}'+a_{--}')$, $\tilde{a}_{-+}'=\frac{1}{\sqrt{2}}(a_{-+}'+a_{--}')$, $\tilde{a}_{--}'=a_{--}'$,
and $\varphi'=\varphi_{+-}'-\varphi_{-+}'$ is the phase difference between $\tilde{a}_{+-}'$ and $\tilde{a}_{-+}'$.
We see now immediately that it is possible to retrieve the rest of the amplitudes.

Measurements on an ensemble of such quantum dot spin systems could be used to observe time-dependent coherent phenomena. For example, the phase $\varphi_{+-}-\varphi_{-+}$ of the two-qubit system initialized in a superposition of singlet and triplet eigenstates would change linearly with time, producing an oscillation in the two-photon interference pattern with a frequency depending on the singlet-triplet energy difference. Observation of a single two-qubit coherent oscillation could also be done by repetitive measurements after initialization (time ensemble). 

For illustration we calculate first the 2-photon correlation for four quantum dots. A general 4-qubit state has the form
$
\left|\psi\right>=
\sum_{\alpha=\uparrow,\downarrow}$$\sum_{\beta=\uparrow,\downarrow}$$\sum_{\gamma=\uparrow,\downarrow}$$\sum_{\nu=\uparrow,\downarrow}$ $a_{\alpha\beta\gamma\nu}\left|\alpha\beta\gamma\nu\right>.
$
$\uparrow$ corresponds to + and $\downarrow$ to - in $a_{\alpha\beta\gamma\nu}$.
After long but straightforward calculation, we obtain
\bea
C_4 & = & C_{1212}+C_{2121}+C_{1313}+C_{3131}+C_{1414}+C_{4141} \nn\\
& & +C_{2323}+C_{3232}+C_{2424}+C_{4242}+C_{3434}+C_{4343} \nn\\
& & +C_{1221}+C_{1331}+C_{1441}+C_{2332}+C_{2442}+C_{3443} \nn\\
& & +C_{1231}+C_{1223}+C_{1241}+C_{1224}+C_{1234}+C_{1243} \nn\\
& & +C_{2113}+C_{2132}+C_{2114}+C_{2142}+C_{2143}+C_{2134} \nn\\
& & +C_{1341}+C_{1332}+C_{1324}+C_{1342}+C_{1334}+C_{3114} \nn\\
& & +C_{3123}+C_{3142}+C_{3124}+C_{3143}+C_{1423}+C_{1432} \nn\\
& & +C_{1442}+C_{1443}+C_{4132}+C_{4123}+C_{4124}+C_{4134} \nn\\
& & +C_{2342}+C_{2334}+C_{3224}+C_{3243}+C_{2443}+C_{4234} \nn\\
& & +C_{1323}+C_{1424}+C_{3132}+C_{4142}+C_{1232}+C_{1434} \nn\\
& & +C_{2123}+C_{4143}+C_{1242}+C_{1343}+C_{2124}+C_{3134} \nn\\
& & +C_{2131}+C_{2434}+C_{1213}+C_{4243}+C_{2141}+C_{2343} \nn\\
& & +C_{1214}+C_{3234}+C_{3141}+C_{3242}+C_{1314}+C_{2324},
\label{C_4}
\eea
where $C_{ijkl}=\sum_{p,q=+,-} |\tilde{a}_{ijpq}||\tilde{a}_{klpq}|\cos(\xi_{ijkl}+\varphi_{ijpq}-\varphi_{klpq})$, 
$\xi_{ijkl}=k(\left|\bx_i-\bx_{\sigma^+}\right|+\left|\bx_j-\bx_{\sigma^-}\right|-\left|\bx_l-\bx_{\sigma^-}\right|-\left|\bx_k-\bx_{\sigma^+}\right|)$, $i,j,k,l=1,2,3$, $i\ne j$, $k\ne l$. $\tilde{a}_{ijpq}$ represents the amplitudes with $\left|\rightarrow\right>_i$ and $\left|\downarrow\right>_j$. The sites $p$ and $q$ can be either $\left|\rightarrow\right>$ or $\left|\downarrow\right>$. An example for $C_4$ is shown in Fig.~\ref{Entanglement_detection_pattern}.
The two-photon correlation
for $n$ quantum dots reads
\bea
C_n & = & \sum_{i=1}^n\sum_{j=1,j\ne i}^nC_{ijij}+\sum_{i=1}^n\sum_{j=1,j>i}^nC_{ijji} \nn\\
& & +\sum_{i=1}^n\sum_{j=1,j\ne i}^n\sum_{k=1,k\ne i}^n\sum_{l=1,l\ne k,l\ne j}^n\hspace{-0.5cm}{'}\hspace{0.5cm}C_{ijkl} \nn\\
& & +\sum_{i=1}^n\sum_{j=1,j\ne i}^n\sum_{k=1,k>i,k\ne j}^n\left(C_{ijkj}+C_{jijk}\right),
\label{C_n}
\eea
where $\sum'$ gives also a vanishing term $C_{ijkl}=0$ if $10i+j>10k+l$ for $i<j$ and $k<l$, $10j+i>10k+l$ for $i>j$ and $k<l$, $10i+j>10l+k$ for $i<j$ and $k>l$, and $10j+i>10l+k$ for $i>j$ and $k>l$.
We obtain
\be
F(n,2)=\left(n\atop 2\right)+\frac{n!}{(n-2)!}\left(n-1\atop 2\right)+2\left(n\atop 2\right)
\ee
$C_{ijkl}$ terms with different spatial frequency dependencies $\cos{\xi_{ijkl}}$. 
The first binomial comes from choosing $i$ and $j$ out of $n$ in the second sum of Eq.~(\ref{C_n}), the second term from
the $\sum'$ sum, and the last binomial from the last sum where $i$ and $k$ are chosen out of $n$.
The number of amplitudes $a$ increases as $2^n$, which is smaller than $F(n,2)$ only for $2<n\le 13$.
In order to increase the number of spatial frequencies $\xi_{ijkl}$, we must use more detectors.
Using unpolarized detectors increases the number of $\xi_{ijkl}$, but does not provide more information on the amplitudes
since the polarized detectors select the amplitudes. Nevertheless, adding unpolarized detectors increases the spatial resolution (see below).
Consequently, we have to add more polarized detectors as $n$ is increased. So we have to measure the 2-, 3-, ..., $N$-photon correlation interferences.
If we use $N$ polarized detectors, we get
\be
F(n,N)=\sum_{j=2}^N\left[3\left(n\atop j\right)+\frac{n!}{(n-j)!}\left(n-1\atop j\right)\right].
\label{F_nN}
\ee
The number of amplitudes $a$ increases as $2^n$, which is smaller than $F(n,N=[n/2])$ $\forall n>2$.
So if we use at least $N=n/2$ detectors and do $N-1$ measurements, we can read out all the amplitudes, provided we have a groundstate or ensemble.
Although the $N=n/2$th summand in Eq.~(\ref{F_nN}) provides already enough equations and frequencies,
only a small part of the amplitudes are visible to the $N=n/2$-photon correlator.
Note that global rotations are needed to read out $a_{+++\cdots +}$ and $a_{---\cdots -}$.

We should also discuss the resolution limit. Since the bandgap of ZnSe is on the order of 500 nm, we have also to use a laser of $\lambda=500$ nm, which means that the qubits must be about 1500 nm apart. From Eq.~(\ref{C_2}) one can see that the resolution can be enhanced by a factor of 2 by varying both detector positions $x_{\sigma^+}$ and $x_{\sigma^-}$ in $\xi_{ijkl}$, such that $\left|\bx_i-\bx_{\sigma^+}\right|-\left|\bx_k-\bx_{\sigma^+}\right|=\left|\bx_j-\bx_{\sigma^-}\right|-\left|\bx_l-\bx_{\sigma^-}\right|$, which leads already to a resolution of $\lambda/2=250$ nm. Further enhancement of resolution can be achieved by measuring the $N$-photon correlations of the form $\left<\bE_{d_N}^{(-)}\ldots\bE_{d_s}^{(-)}\bE_{\sigma^+}^{(-)}\bE_{d_r}^{(-)}\ldots\bE_{d_q}^{(-)}\bE_{\sigma^-}^{(-)}\bE_{d_p}^{(-)}\ldots\bE_{d_1}^{(-)}\right.$\\$\left.\bE_{d_1}^{(+)}\ldots\bE_{d_p}^{(+)}\bE_{\sigma^-}^{(+)}\bE_{d_q}^{(+)}\ldots\bE_{d_r}^{(+)}\bE_{\sigma^+}^{(+)}\bE_{d_s}^{(+)}\ldots\bE_{d_N}^{(+)}\right>$, which enhances the resolution by a factor of $N$ if $\left|\bx_g-\bx_{d_1}\right|-\left|\bx_w-\bx_{d_1}\right|=\ldots=\left|\bx_i-\bx_{\sigma^+}\right|-\left|\bx_k-\bx_{\sigma^+}\right|=\left|\bx_j-\bx_{\sigma^-}\right|-\left|\bx_l-\bx_{\sigma^-}\right|=\ldots =\left|\bx_v-\bx_{d_N}\right|-\left|\bx_h-\bx_{d_N}\right|$ for all pairs of distances between quantum dots and detectors. Since $N\le n$, the larger $n$, the higher the resolution that can be achieved. However, as soon as the entanglement decays, also the resolution decreases. Nevertheless, this decrease in resolution means that the spatial oscillation period $nk$ decreases down to $k$, which can be measured in real-time. For present quantum dot technology $N=10$ would be suitable. Another solution would be to use only every e.g. fifth quantum dot as a computational qubit, and the other quantum dots would be used to mediate a super-superexchange interaction between the computational qubits, which requires a generalization of the superexchange mechanism presented in Ref.~\cite{Recher}. Using three intermediate dots gives an exchange coupling $J\sim t_0^8/\epsilon^7$, where $t_0$ is the tunnel coupling between the dots, and $\epsilon$ is the energy difference between the level of the computational dots and the level of the intermediate dots. $J$ can be maximized by increasing both $t_0$ and $\epsilon$ while keeping $t_0\lesssim\epsilon$\cite{Leuenberger2001}.
For a test of our theory, an appropriate experimental system would be similar to the one found in \cite{Ouyang}.
Singlet-triplet oscillations could also be observed in an ensemble of paired coupled dots of different sizes.
In addition, it would be interesting to observe the quantum Zeno effect induced by repetitive measurements on the singlet and triplet states.
Finally, it is also possible to measure the $p$-qubit entanglement of an $n$-qubit system, where $p<n$, by focusing the laser beams on only the $p$ quantum dots.

{\it Acknowledgement}. We thank G.~Burkard, R.~Epstein, O.~Gywat, and M.~Ouyang for useful discussions. We acknowledge the support of DARPA/ARO, DARPA/ONR, and the US NSF.

\end{document}